\documentclass[
aip,
 amsmath,amssymb,
 reprint,%
]{revtex4-1}

\usepackage{graphicx}% Include figure files
\usepackage{dcolumn}% Align table columns on decimal point
\usepackage{bm}% bold math

\usepackage[mathlines]{lineno}% Enable numbering of text and display math
\usepackage[utf8]{inputenc}
\usepackage[T1]{fontenc}
\usepackage{mathptmx,etoolbox,textcomp,gensymb,upgreek}
\usepackage[dvipsnames]{xcolor}
\usepackage[normalem]{ulem}

\newcommand{\um}{$\upmu$m}
\newcommand{\uA}{$\upmu$A}
\newcommand{\us}{$\upmu$s}
\newcommand{\Tc}{$T_C$}
\newcommand{\Top}{$T_{op}$}
\newcommand{\Tsc}{0.39 $T_C$}
\newcommand{\Tcal}{0.96 $T_C$}
\newcommand{\Isw}{$I_{sw}(T)$}
\newcommand{\IswSC}{$I_{sw}^{Geiger}$}
\newcommand{\IswCal}{$I_{sw}^{Cal}$}
\newcommand{\Ibias}{$I_{bias}$}
\newcommand{\Power}{$E_{abs}$}
\newcommand{\Amp}{$A_{sig}$}
\newcommand{\AmpM}{$\overline{A}_{sig}$}
\newcommand{\tauf}{$\tau_{\textit{fall}}$}
\newcommand{\taufSC}{$\tau_{\textit{fall}}^{Geiger}$}
\newcommand{\taufCal}{$\tau_{\textit{fall}}^{Cal}$}
\newcommand{\Lk}{$L_{k}$}
\newcommand{\Sup}{$Supplementary$ $Material$}

\makeatletter
\def\@email#1#2{%
 \endgroup
 \patchcmd{\titleblock@produce}
  {\frontmatter@RRAPformat}
  {\frontmatter@RRAPformat{\produce@RRAP{*#1\href{mailto:#2}{#2}}}\frontmatter@RRAPformat}
  {}{}
}%
\makeatother
\begin{document}

\preprint{AIP/123-QED}

\title[Dual-Mode Calorimetric Superconducting Nanowire Single Photon Detectors]{}
% Authors below: \\
\author{Hsin-Yeh Wu}
 \email{wuhsinyeh@ntu.edu.tw}
\affiliation{ 
Department of Physics, National Taiwan University, Taipei 10617, Taiwan}%
\author{Marc Besançon}
\affiliation{%
CEA-IRFU, Paris-Saclay University 91190 Gif sur Yvette cedex, France}%
\author{Jia-Wern Chen}
\affiliation{Research Center for Applied Sciences, Academia Sinica, Taipei 11529, Taiwan}%
\author{Pisin Chen}
\affiliation{ 
Department of Physics, National Taiwan University, Taipei 10617, Taiwan}%
\affiliation{Leung Center for Cosmology and Particle Astrophysics, 
National Taiwan University, Taipei 10617, Taiwan}
\author{Jean-François Glicenstein}
\affiliation{%
CEA-IRFU, Paris-Saclay University 91190 Gif sur Yvette cedex, France}%
\author{Shu-Xiao Liu}
\affiliation{ 
Department of Physics, National Taiwan University, Taipei 10617, Taiwan}%
\author{Yu-Jung Lu}
\affiliation{ 
Department of Physics, National Taiwan University, Taipei 10617, Taiwan}%
\affiliation{Research Center for Applied Sciences, Academia Sinica, Taipei 11529, Taiwan}%
\author{Xavier-François Navick}
\affiliation{%
CEA-IRFU, Paris-Saclay University 91190 Gif sur Yvette cedex, France}%
\author{Stathes Paganis}%
 \email{paganis@phys.ntu.edu.tw}
 \affiliation{ 
Department of Physics, National Taiwan University, Taipei 10617, Taiwan}%
\affiliation{Leung Center for Cosmology and Particle Astrophysics, 
National Taiwan University, Taipei 10617, Taiwan}
\author{Boris Tuchming}
\affiliation{%
CEA-IRFU, Paris-Saclay University 91190 Gif sur Yvette cedex, France}%
\author{Dimitra Tsionou}
\affiliation{ 
Department of Physics, National Taiwan University, Taipei 10617, Taiwan}%
\author{Feng-Yang Tsai}
\affiliation{ 
Department of Physics, National Taiwan University, Taipei 10617, Taiwan}%
\affiliation{Research Center for Applied Sciences, Academia Sinica, Taipei 11529, Taiwan}%

\date{\today}% It is always \today, today,

\begin{abstract}
A dual-operation mode SNSPD is proposed. In the conventional Geiger mode, the sensor operates at temperatures well below the critical temperature, \Tc, working as an event counter without sensitivity to the number of photons impinging the sensor.
In the calorimetric mode, the detector is operated at temperatures just below \Tc~and displays calorimetric sensitivity in the range of 15 to 250 absorbed-photon energy equivalent for a photon beam with a wavelength of 515~nm.
In this energy sensitive mode, photon absorption causes Joule heating of the SNSPD that becomes partially resistive without the presence of latching. Depending on the application,
by tuning the sample temperature and bias current using the same readout system, the SNSPD can readily switch between the two modes.
In the calorimetric mode, SNSPD recovery times shorter than the ones in the Geiger mode are observed, reaching values as low as 560~ps. 
%\textcolor{red}{Although the ultimate goal of this work is to demonstrate dual-mode SNSPDs with single-photon sensitivity, this has not yet been achieved due to latching issues (Geiger mode) and noise limitations in our setup (calorimetric mode).}
Dual-mode SNSPDs may provide significant advancements in spectroscopy and calorimetry, where precise timing, photon counting and energy resolution are required.

\end{abstract}

\maketitle

\section{\label{sec:intro}Introduction}

During the past three decades, superconducting photon detector development has been advancing at a rapid pace. This progress is largely due to their low intrinsic energy threshold of order meV and dark-count rate (DCR) or noise equivalent power compared to traditional semiconductor photon detectors. Superconducting nanowire single-photon detectors (SNSPDs) are state-of-the-art single-photon sensors that have been widely adopted in various photonic applications. These applications include quantum key distribution (QKD)~\cite{beutel2021}, quantum information~\cite{you2020}, detection of luminescence from singlet oxygen~\cite{gemmell2013}, LiDAR applications~\cite{taylor2019} and more. SNSPDs exhibit several outstanding characteristics: near 100\% system detection efficiency~\cite{reddy2020,chang2021}, excellent timing resolution with low timing jitter~\cite{chang2021}, extremely low DCR, very high count rate, and a broad spectral operating range~\cite{esmaeilzadeh2021}.

Conventional SNSPD operation temperatures, \Top, are well below the superconducting critical temperature, \Tc. The detectors are biased with a direct bias current, \Ibias, just below the switching current (defined as the current at which the SNSPD starts becoming resistive), \Isw. 
Absorption of photons induce local hotspots in the nanowire that extend in the full width of the wire, causing the bias current density to exceed the critical current density and trigger a transition from the superconducting state to the normal-conducting state. This process enables SNSPDs to operate in a single-photon counting mode~\cite{engel2015}.  While SNSPDs are highly effective in the single-photon counting (Geiger) mode, they are not sensitive to the number of photons impinging the detector, or equivalently, they are not energy sensitive. There have been attempts to fabricate photon-number-resolving SNSPDs, where multiple photons create multiple local hotspots, inducing different discrete resistance changes~\cite{cahall2017, zhu2020}. However, the dynamic range of these devices remains limited.

In contrast, other types of superconducting detectors, such as transition-edge sensors (TES) and microwave kinetic inductance detectors (MKID), offer microcalorimeter capabilities. These devices provide good energy resolution by measuring the temperature rise or changes in the surface impedance of a superconductor through the kinetic inductance effect induced by photon absorption. 
%This allows for precise energy measurements. 
However, despite their precise photon energy measurement capability, TES and MKID detectors typically exhibit slower response times compared to SNSPDs~\cite{eisaman2011,swimmer2023}. This trade-off between energy resolution and timing capability is a critical consideration when selecting the appropriate detector for specific applications.

\begin{figure*}
    \includegraphics[width=.33\textwidth]{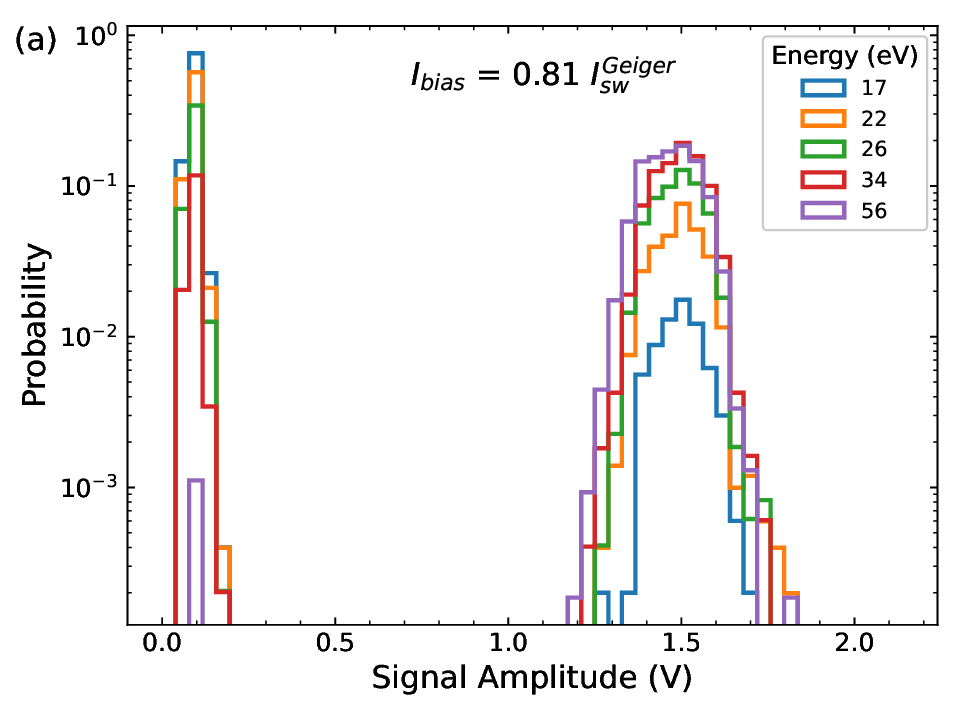}
    \includegraphics[width=.33\textwidth]{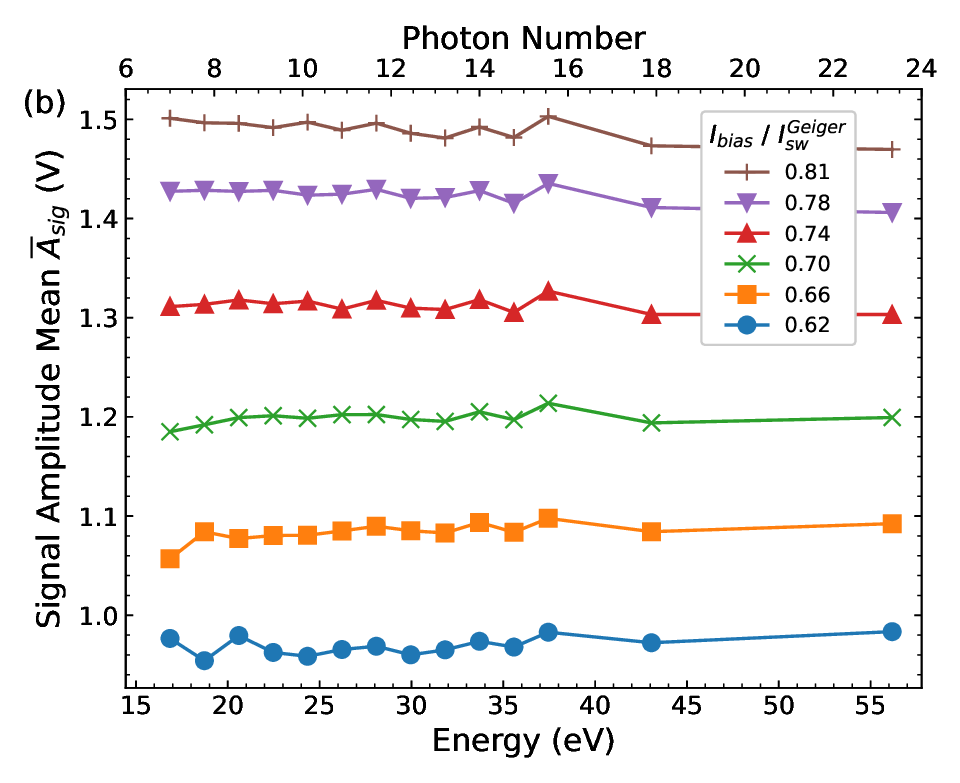}
    \includegraphics[width=.33\textwidth]{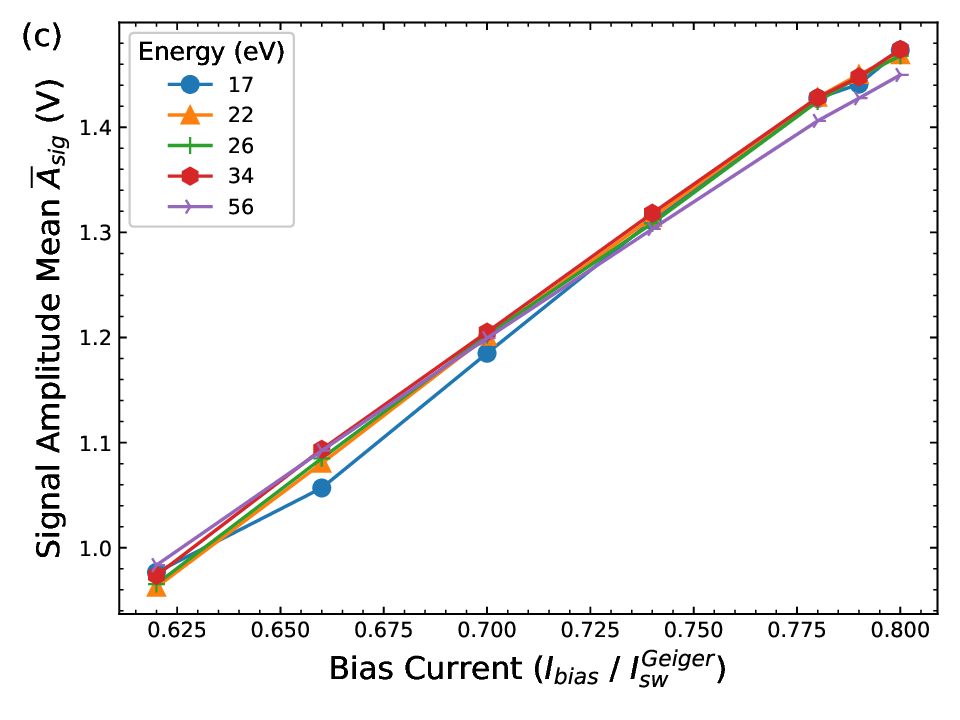}
\caption{Geiger mode: The SNSPD is operated at 4.68K (\Tsc). (a) Recorded signal amplitude distributions for different absorbed energies normalized to unity, for a bias current set at 0.81  \IswSC. (b) Signal amplitude mean as a function of the absorbed energy, with the equivalent absorbed photon number indicated on the top axis, for a range of bias currents. (c) Signal amplitude mean as a function of the bias current, for a range of absorbed energies. The lines connecting the data points serve as visual guides to illustrate the trends.
\label{fig:1}
}
\end{figure*}
Several newly proposed experiments that employ single-photon sensors, require energy resolution and stringent timing capabilities, including precise timing resolution and minimal timing jitter, with a very low DCR. For instance, the AnaBHEL (Analog Black Hole Evaporation via Lasers) experiment~\cite{chen2022a} aims to detect entangled photon pairs from Hawking-analogue radiation produced by accelerating plasma mirrors. This experiment demands very low timing jitter and high timing resolution to distinguish the entangled photon pairs from the irreducible photon background generated by the laser plasma. Good energy resolution is also required to validate the blackbody radiation energy spectrum of the Hawking photons. Another relevant research initiative is the next generation of neutrinoless double-beta decay experiments, such as CUPID (CUORE Upgrade with Particle IDentification)~\cite{cupid}, where one needs to resolve an irreducible background to two-neutrino double-beta decay (2$\nu\beta\beta$) events from scintillating crystals. These background events contribute to the experiment as pileup, necessitating excellent timing resolution in order to resolve them. Furthermore, quantum information and quantum key distribution experiments, aiming at high repetition rates exceeding GHz~\cite{anderson2020,beutel2021}, impose stringent timing requirements on the photon detectors.

%\textcolor{red}{In this work, we investigate the dual-mode operation of an SNSPD in a high photon flux regime. 
%Measurements on prototypes exposed to pulsed-laser beams
%with thousands of photons per laser pulse, display a sensitivity in the range of 15-250 absorbed-photon energy equivalent given our device's 1\% photon absorption rate. The detector's design closely aligns with typical SNSPDs, incorporating similar geometry, superconducting material selection, and readout techniques. However, our investigation extends beyond traditional single-photon detection to explore the device's capabilities in multi-photon detection.}

The conventional SNSPD single-photon operation mode (the Geiger mode) leads to pulse amplitudes that are constant, independent of the number of photons impinging the sensor.
The widely accepted physics mechanism behind this behaviour
is due to vortex crossing from one side of the wire to the other and vortex-vortex annihilation in the center of the wire~\cite{Bulaevskii_2011}. 
In this mode the wire is biased at relatively high bias current of order 100~$\mu$A increasing the probability of vortex-crossing.
After photon absorption, the formation of a hot spot reduces the potential barrier for vortex crossing and leads to the rapid growth of the hot spot region to the entire width of the wire.  This process has been shown~\cite{Bulaevskii_2011} to result to pulse amplitudes independent of the number of photons absorbed by the sensor. 

Extending beyond the conventional Geiger-mode operation, we explore a high-speed SNSPD calorimetric mode, 
where the detector 
displays calorimetric sensitivity in the range of 15-250 absorbed photons for a photon wavelength of 515~nm.
%By calorimetric, we mean sensitivity 
When the SNSPD is operated just below  \Tc, the switching current \IswCal{}~is an order of magnitude lower than the one in the Geiger mode. Operating the detector at a bias current close to  \IswCal{}~ leads to suppression of vortex crossing. Absorption of photons does not induce vortex crossing but instead leads to Joule heating of the SNSPD that becomes partially resistive and thus the amplitude of the observed pulses increases with the number of absorbed photons. 

In the following sections we demonstrate the dual-\-operation mode of SNSPDs: 
(i) a {\it Geiger} single-photon or multi-photon event counting mode and 
(ii) an energy sensitive {\it calorimetric} mode.
 This dual-mode capability allows the SNSPD to switch between high-precision event counting and energy measurement by adjustment of \Ibias ~and \Top, thus offering a versatile solution for high-speed spectroscopy applications.
%There is a lack of high-speed calorimeter or spectrometer, a challenge that could potentially be addressed by the superconducting calorimeter proposed in this work. We explore the use of an SNSPD as a calorimeter by controlling the sample temperature of a NbN SNSPD to demonstrate two distinct modes of signal response. The first mode, Geiger mode, occurs when the sample operates well below the critical temperature (\Tc). The second mode, calorimetric mode, is achieved when the sample operates close to the critical temperature.

\section{\label{sec:setup}SNSPD Fabrication and Experimental Setup}
The first step of the fabrication process of the SNSPDs used in this work, is the deposition of 15~nm thin NbN films on MgO substrates using RF sputtering under the following conditions: 120 W power, 36:0.1 sccm Ar:N$_2$ flow rate, 0.9 mTorr pressure, and a substrate temperature of 800°C. For more detailed information on the NbN film characterization, the reader should refer to Ref.~\onlinecite{yang2023} and its supporting documentation. The films are subsequently patterned into a nanowire meander structure using electron-beam lithography (EBL) with ZEP520A resist mixed with ZEP-A at a 1:1 ratio, and developed with ZED-N50 developer. The nanowire meander has a width of 200~nm, a pitch of 350~nm, and a total length of 400~\um, covering a sensing area of 12~\um{} by 12~\um. The structure is etched using reactive ion etching with a CF$_4$:O$_2$ flow rate of 20:2~sccm and a DC power of 200~W. The measured critical temperature (\Tc) of the SNSPD is 12~K.

During characterization and testing, 
the SNSPD is mounted on the 4K stage of a pulse-tube cryocooler with custom-made packaging. In this work, we demonstrate the two distinct operating modes by adopting two operating temperatures: The first is the Geiger mode at 4.68~K (\Tsc), characterized by a switching current, \IswSC, of 127~\uA{} and a superconducting transition width of 0.2~\uA, indicating a sharp superconducting transition. The transition width is defined as the difference between the current at which the detector becomes fully resistive at 335~k$\Omega$ and the switching current, \Isw. The second is the calorimetric mode operated at 11.47~K (\Tcal), with a switching current, \IswCal, of 13~\uA{} and a superconducting transition width of 8~\uA, indicating a slow and wide superconducting transition. In this mode, the SNSPD can operate both in the superconducting state with \Ibias{} below \IswCal{}, or in a partially resistive state with \Ibias{} above \IswCal{} before fully transitioning to the non-superconducting state. The reader should refer to \Sup{} for more information on the resistance-current diagram of this sample.

During SNSPD photoresponse measurements, we employ a picosecond pulsed laser (Prima from PicoQuant) featuring three visible wavelengths, a pulse width of less than 100~ps, and a variable repetition rate ranging from 1~kHz to 200~MHz. We select a 515~nm wavelength as our primary source. The laser is coupled to the SNSPD sensing area through open air, with the laser spot size focused to a diameter of 3~\um{}. The SNSPD is biased through a bias tee (Minicircuits ZFBT-4R2GW+), and output signal pulses are amplified using two room-temperature low-noise amplifiers (RFBay LNA-1030, with a bandwidth of 20~MHz to 1~GHz, 30~dB gain, and a noise figure of 1.3 dB). The amplified signal is digitized by a PXI Express Oscilloscope with an external trigger from the pulsed laser (NI PXIe-5162, featuring 1.5~GHz bandwidth, 2.5~GS/s per channel sampling rate, and an effective number of bits of 7.0 at 50$\Omega$). A data acquisition gate of 1000 time samples (400~ns) is set for each laser trigger, with the trigger reference position at 30\%. This setup, typically used for a pulsed-laser repetition rate of 10~kHz, covers a single signal pulse with a duration of approximately 40~ns and a measured latency of 20~ns 
with respect to the trigger time of arrival. 

We define as laser energy the expected energy per laser pulse after accounting for the measured attenuation from the point of emission to the sensor surface. Absorbed energy, denoted as \Power, is the laser energy adjusted for a 1\% photon absorption efficiency. This absorption estimation is detailed in the Discussion section. The detailed experimental setup and example signal-pulse spectra are presented in the \Sup.

\begin{figure*}
    \includegraphics[width=.33\textwidth]{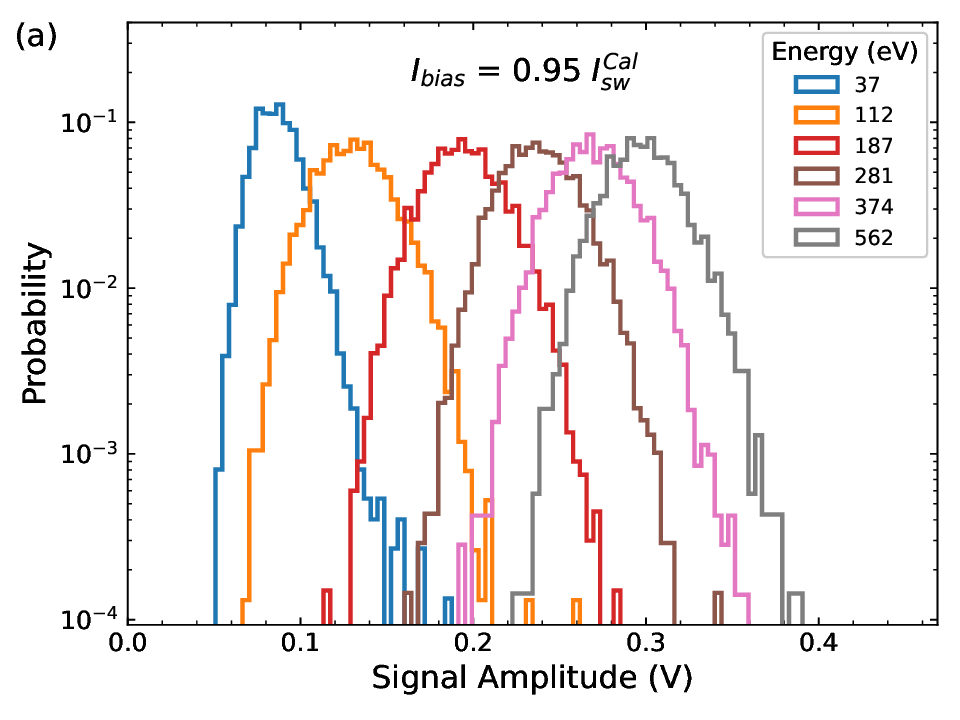}
    \includegraphics[width=.33\textwidth]{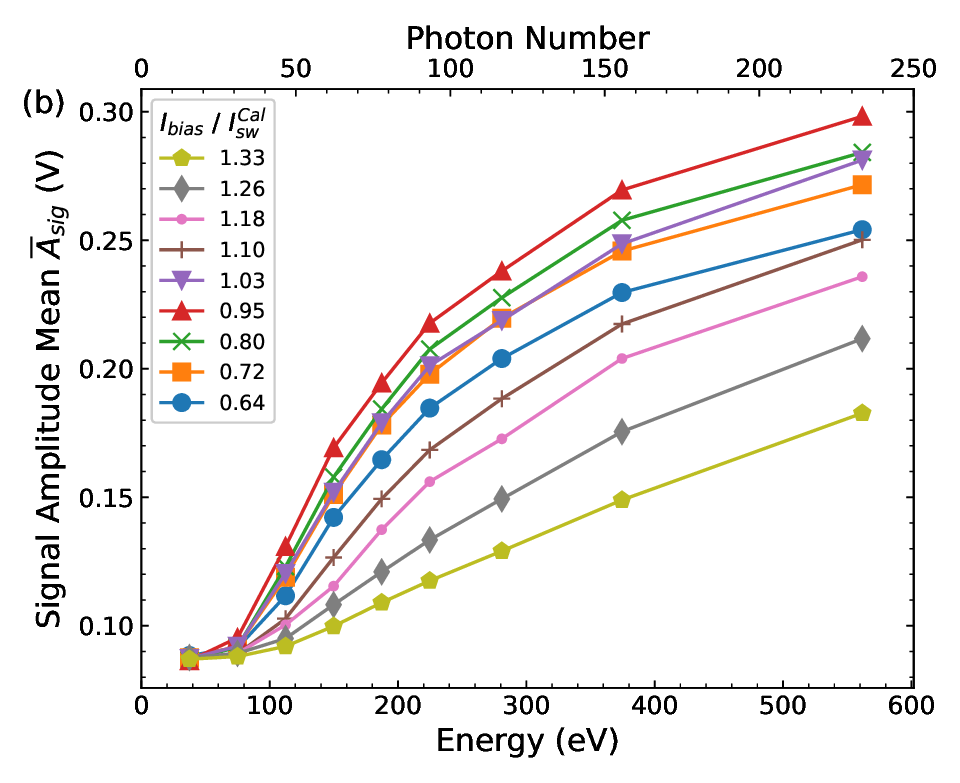}
    \includegraphics[width=.33\textwidth]{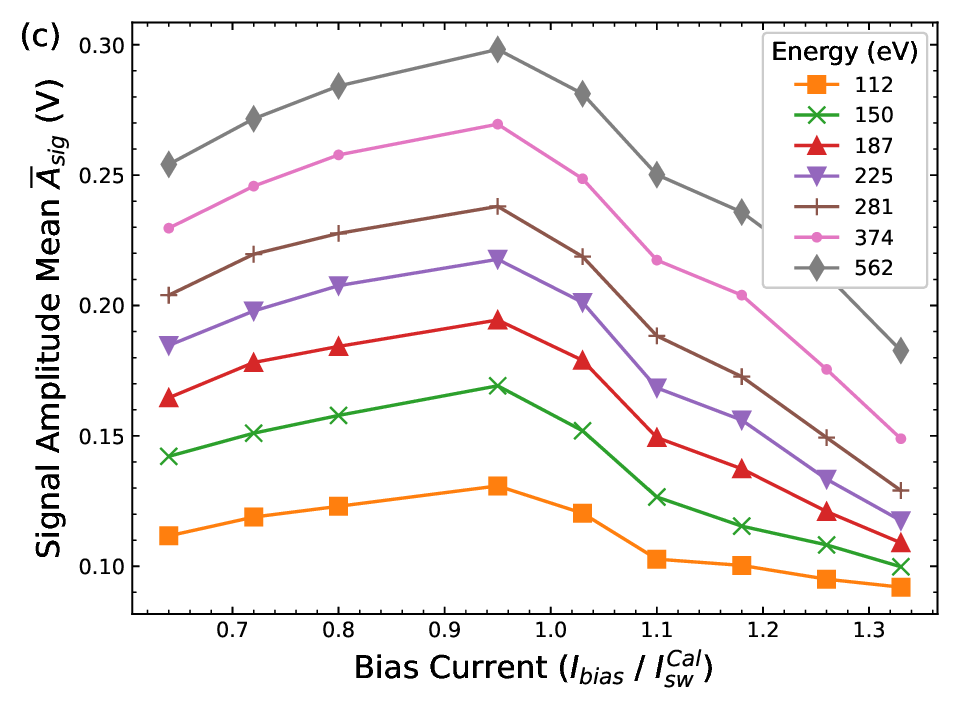}
\caption{Calorimetric Mode: The SNSPD is operated at 11.47K (\Tcal). (a) 
Recorded signal amplitude distributions for different absorbed energies normalized to unity, for a bias current set at 0.95 \IswCal. The amplitudes show a clear energy dependence. (b) 
Signal amplitude mean as a function of the absorbed energy, with the equivalent absorbed photon number indicated on the top axis, for a range of bias currents.
(c) Signal amplitude mean as a function of the bias current, for a range of absorbed energies. The lines serve as visual guides to illustrate the trends.
\label{fig:2}
}
\end{figure*}

\section{\label{sec:results}RESULTS}
In the following, we present the results obtained with the SNSPD operating at two different temperatures. First, for \Tsc{} in the Geiger mode, the dependence of the signal pulse amplitude, \Amp, on the expected energy per laser pulse \Power{} and \Ibias{} is shown in Fig.~\ref{fig:1}. 
The amplitude \Amp{} of each laser pulse is measured offline and defined as the pulse peak-to-peak voltage difference within a fixed 25~ns time gate. This gate starts 1~ns before the triggered signal-pulse arrival, to approximately 10~ns after the lowest level of the overshoot from a typical signal pulse as shown in the \Sup. The timing jitter of the signal pulses is shorter than the sampling rate, leading to a negligible probability of the fixed gate to miss the signal.

Fig.~\ref{fig:1}~(a) shows the amplitude, \Amp{}, distributions (normalized to unity) at different absorbed energies for a \Ibias{} of 0.81~\IswSC. To extract these amplitudes for each energy, 10,000 pulsed-laser triggers are used. For each energy there are two distinct groups of events: the signal events, peaked at 1.49 V, representing the fraction of absorbed-photon signal pulses that coincide with the laser trigger, and the noise events, peaked at 95 mV, representing the fraction of events producing no SNSPD signal. In this case, \Amp{} corresponds to the peak-to-peak amplitude of the electronic noise. As \Power{} increases, the fraction of the signal events increases, while the average amplitude of the signal distribution is constant, indicating that \Power{} correlates to the laser pulse detection efficiency but not to the signal amplitude mean, \AmpM. \AmpM ~is obtained by a Gaussian fit of the signal peak. Fig.~\ref{fig:1}~(b) displays the dependence of the signal amplitude mean to the absorbed energy, \Power{}, for different settings of the bias current, \Ibias{}. As expected for an SNSPD operating in the Geiger mode for fixed bias current, no dependence of the signal amplitude on the absorbed energy is observed.
%For all the provided \Ibias{}, \AmpM{} did not depend on %\Power, indicating a Geiger-mode operation. 
In Fig.~\ref{fig:1}~(c) the dependence of the amplitude mean to the normalized bias current for fixed \Power{} is shown.
%Switching to a profile of \Ibias{} at different fixed \Power{}, 
A linear correlation between \AmpM{} and \Ibias{} is observed. %For bias currents higher than 0.81~\IswSC{}, there is a latching 
%effect.

In the calorimetric mode, where the SNSPD is operated at \Tcal{}, the signal amplitude distribution and means are shown in Fig.~\ref{fig:2}. In Fig.~\ref{fig:2}~(a), the amplitude distributions (normalized to unity) are shown for a bias current \Ibias{} at 0.95~\IswCal~and different absorbed energies. In this mode we observe that all events recorded are signal events
and as the energy increases, the Gaussian-shape amplitude distributions shift to higher values.
%there was only a single distribution instead of two distinct %distributions as shown in Fig.~\ref{fig:1}~(a) for each 
%This suggests that there was no clear separation between a %detected laser pulse and a lost one. Secondly, the signal %distribution shifted increasingly with increased \Power, %indicating that \Power{} correlates with \AmpM. 
%These observations suggest that the SNSPD was not operating %in Geiger-mode but rather in a calorimetric mode.
\begin{figure*}
    \includegraphics[width=.33\textwidth]{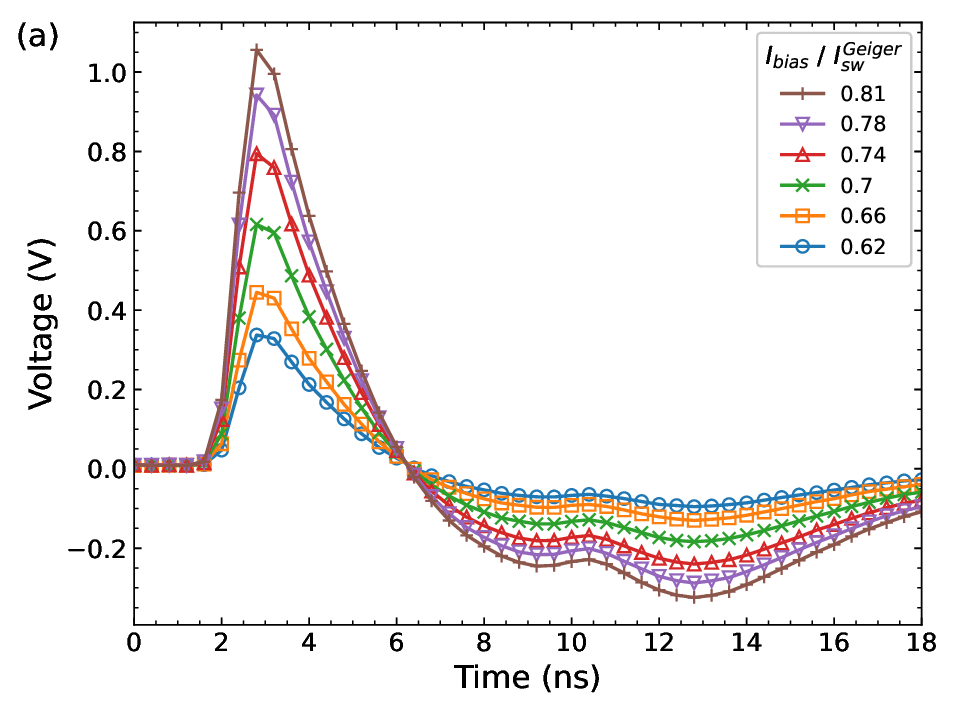}
    \includegraphics[width=.33\textwidth]{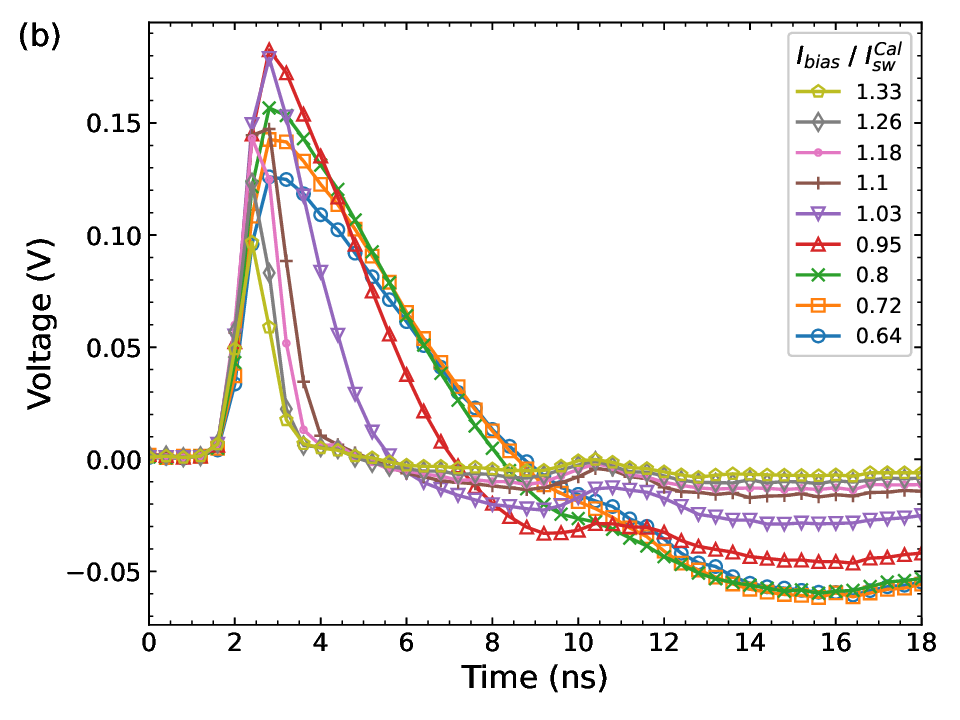}
    \includegraphics[width=.33\textwidth]{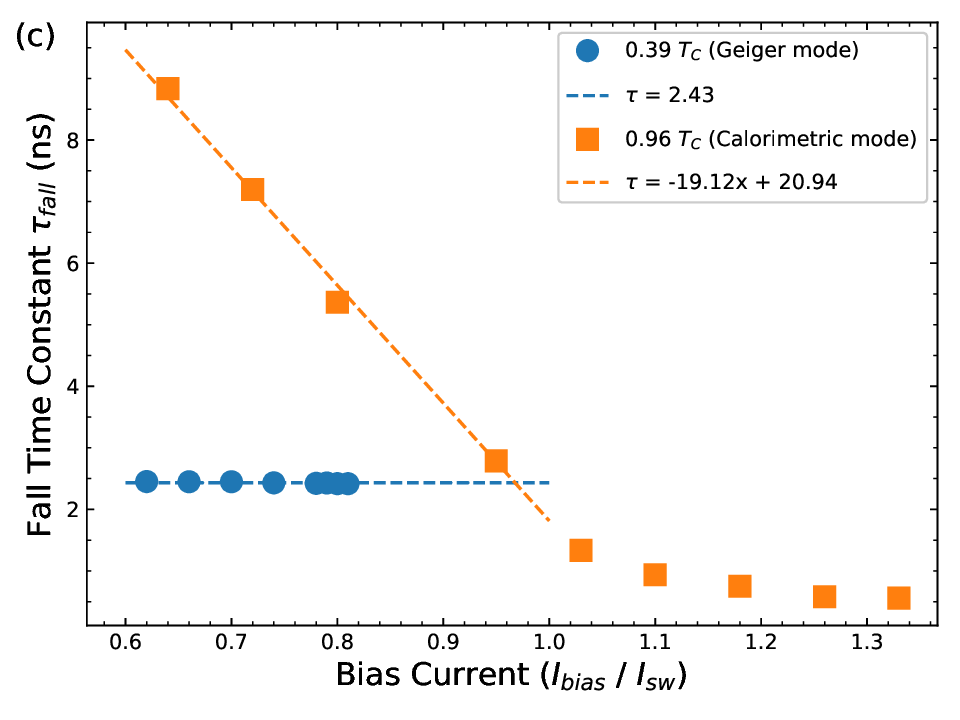}
\caption{Sample SNSPD signal spectra, averaged over 1000 laser events at a fixed \Power{} for a range of bias currents, for the Geiger mode (a) and calorimetric mode (b). In these spectra, the markers represent the digitized samples, while the lines serve as a visual guide. An undershoot in the spectrum is evident due to the bandwidth limitations of the low-noise amplifiers. (c) The fitted-pulse falling time constant (\tauf) is presented as a function of the normalized bias current \Ibias/\Isw{} for both Geiger and calorimetric modes. For the Geiger mode, the data are shown up to 0.81~\IswSC, as the SNSPD latches above this bias current. For data points corresponding to bias currents smaller than the switching current, \Isw, the results of two linear fits are also shown.
\label{fig:3}
}
\end{figure*}
The dependence of the amplitude mean, \AmpM{}, on \Power{} for different bias currents, is presented in Fig.~\ref{fig:2}~(b). 
%The detector gain, $G$, defined as \AmpM{}/\Power, %characterizes the signal amplitude for a given energy input. %\AmpM{} increases non-linearly with \Power{} and saturates %at high \Power. Additionally, $G$ 
The amplitude mean increases non-linearly with \Power, with the increase being slower at high absorbed energies.
The amplitude is maximum for bias currents just below the switching current.
%varies with \Ibias{}—it increases with \Ibias{} and reaches %a maximum at the switching current before decreasing as %\Ibias{} exceeds the critical current. Thus, $G$ can be %expressed as $G(E_{Laser}, I_{bias})$. 
At lower energy levels, there is a region where \AmpM{} 
is almost constant, a result of the SNSPD signal amplitude being smaller than the electronic noise and \AmpM{} becoming noise-dominated. 
The dependence of the signal amplitude mean on the bias current 
for different absorbed energies, is shown in Fig.~\ref{fig:2}~(c). For all energies, an initial increase of the amplitude with \Ibias{} is observed until the bias current reaches the switching current. Beyond this point, \AmpM{} begins to decrease, due to the partial transition of the SNSPD from the superconducting to the resistive state.

%Furthermore, the SNSPD can be operated above \IswCal{}, where it is initially in a resistive state. This behavior is more clearly illustrated in the \Ibias{} profile shown in Fig.\ref{fig:2}~(c). For a fixed \Power{}, \AmpM{} initially increases with \Ibias{} until it approaches \IswCal{}. Beyond this point, \AmpM{} begins to decrease, demonstrating the transition dynamics of the SNSPD from the superconducting to the resistive state.

In addition to the pulse amplitude, a pulse-shape analysis is performed. The pulse shape also exhibits differences when comparing the Geiger with the calorimetric mode. Figures \ref{fig:3}~(a) and \ref{fig:3}~(b) display the pulse shapes, averaged over 1000 signal pulses to mitigate electric noise contributions. The averaged shapes are shown for different \Ibias{} values at a fixed \Power{} for the two operation modes. The pulse falling time constant, \tauf, shown in Fig.~\ref{fig:3}~(c), indicates the time the bias current needs to return from the load resistance to the SNSPD. The load resistance, $R_{\text{L}}$, the SNSPD resistance, $R_{\text{SNSPD}}$, and the SNSPD kinetic inductance, \Lk, are related by the formula: $\tau_{\textit{fall}} = L_{\text{k}} / (R_{\text{L}} + R_{\text{SNSPD}})$. \tauf{} is obtained by performing an exponential fit to the falling edge of the pulse, using the expression: $Ae^{(-(t - t_{0}) / \tau_{\textit{fall}})} + C$. In the Geiger mode, the signal pulse shapes exhibit an amplitude increasing with the bias current, while the superconducting state pulse falling time, \taufSC{}, remains approximately constant at a value of 2.43~ns.
%increased \Amp{} with an approximately constant \taufSC{} of %2.43 ns as \Ibias{} increased. 
For this constant \taufSC{} and $R_{\text{L}}=50~\Omega$, $R_{\text{SNSPD}}=0~\Omega$, a kinetic inductance \Lk{} of 121.5 nH is extracted. 
In the calorimetric mode, the falling time, \taufCal{}, is longer than 
the constant \taufSC{} for bias currents below the switching current \IswCal. It decreases linearly with increasing bias current, with a rate of 1.47 ns/\uA{}, and becomes smaller than \taufSC{} above \IswCal.
From these measurements we observe the kinetic inductance decreasing
with a rate of 73.5 nH/\uA{}. 
For larger values of bias current, the time constant decreases non-linearly approaching a saturation value of approximately 560~ps.
As $R_{\text{SNSPD}}$ is no longer zero but increasing non-linearly with \Ibias{}, the saturated \taufCal{} indicates a non-linear increase in \Lk. 
Notably, operating with \Ibias{} above \IswCal{} provides a faster reset time compared to the conventional Geiger mode. As detailed in the \Sup{} where plots of \tauf{} at different \Power{} in both modes are presented, \tauf{} does not show strong dependence on \Power{}.

Finally, it is important to note that while operating in the calorimetric mode, the DCR was measured using self-triggering and was found to be negligible.

\section{\label{sec:Discussion}Discussion}

From the results presented in the previous section, it is clear that in the calorimetric mode the measured signal is due to a different physical mechanism than in the case of the conventional Geiger mode. In both modes, the absorbed photons trigger a sudden increase in the SNSPD resistance that redirects the bias current to the load resistance, resulting in an output signal pulse~\cite{yang2007}. In the Geiger mode, the SNSPD resistance transitions from zero to a large constant value corresponding to the resistance of the normal conducting state. The measured resistance of the SNSPD conducting state is approximately 335~k$\Omega$. Since this transition leads to a constant jump of the SNSPD resistance, a constant signal amplitude for a fixed \Ibias~is observed. Variations in \Ibias~lead to a linear change in signal amplitude, consistent with the results presented in Fig.~\ref{fig:1}.

\begin{figure}[h!tbp]
\centering
    \includegraphics[width=.48\textwidth]{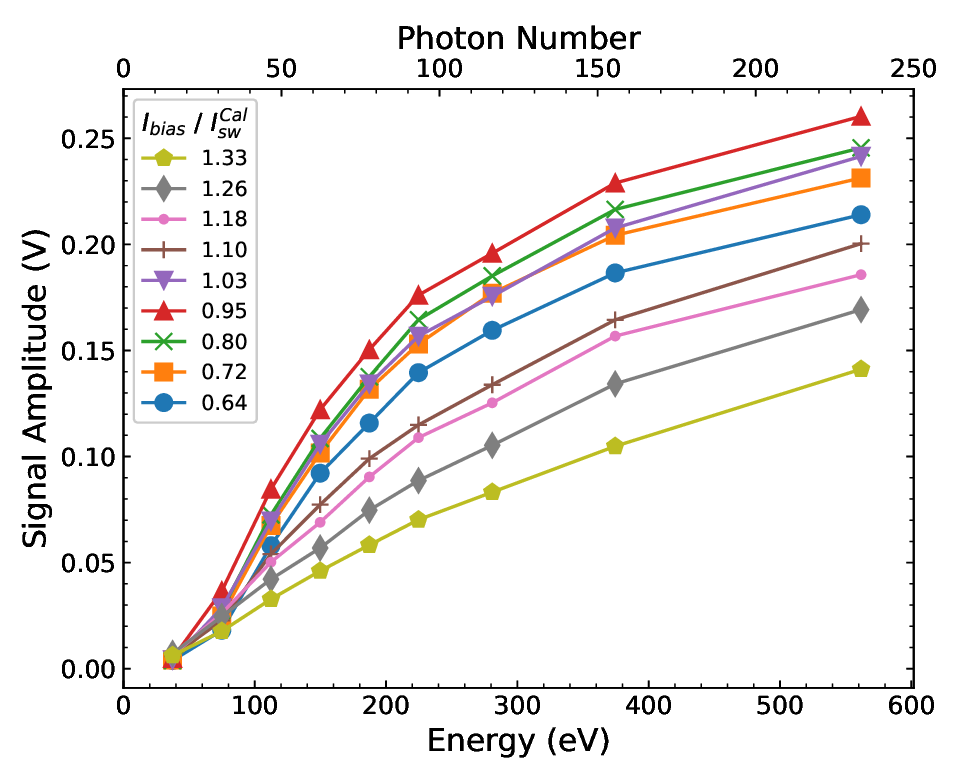}
\caption{Signal pulse amplitude mean, \AmpM{}, 
as a function of absorbed energy in the calorimetric mode, with the equivalent absorbed photon number indicated on the top axis. Unlike Fig.~\ref{fig:2}~(b), where \AmpM{} is measured directly from individual laser events, here \AmpM{} is obtained by averaging over 1000 pulsed-laser events to reduce electronic noise, and subsequently measuring the peak-to-peak voltage range of the averaged pulse spectrum. This approach 
mitigates the inefficiency at low \Power{} due to a poor signal-to-noise ratio, thereby improving the behaviour of the amplitude mean observed at low \Power{} in Fig.~\ref{fig:2}~(b). The lines connecting the data points in the plot serve as visual guides.
\label{fig:4}
}
\end{figure}

In the calorimetric mode the behaviour of the signal amplitude and the pulse shape suggest a Joule heating mechanism in action. This mechanism dominates the vortex-crossing mechanism that is considered to be responsible for the Geiger operation mode. One reason for the suppression of vortex crossing is due to the bias current in the calorimetric mode being an order of magnitude lower than that in the Geiger mode.
Assuming the same electrical model for both modes, when operating in the calorimetric mode, after the absorption of a certain amount of energy the nanowire experiences a partial transition to the resistive state. 
%This is because in this mode the operation temperature is just below the measured Curie temperature and for a particular range of number of photons absorbed, the sensor only becomes partially resistive. 
The absorbed photons cause Joule heating in the SNSPD leading to an increase in the SNSPD resistance. 
At high absorbed energies or equivalently high photon numbers absorbed, the SNSPD is expected to become fully resistive and approach a resistance of 335~k$\Omega$. As shown in Fig.~\ref{fig:4}, our measurements show a signal amplitude that saturates at a value consistent with the normal-state resistance.
It is important to note again that the typical bias currents in the calorimetric mode are significantly lower than the ones in the Geiger mode, thus the corresponding signal pulse amplitudes are also lower.
If the SNSPD is operated with a bias current above \IswCal, 
the signal amplitude decreases, as expected, since the initial state of the SNSPD is already partially resistive.

The SNSPD calorimetric sensitivity to very low numbers of absorbed photons 
is 
%The sensitivity of the SNSPD to a low overall photon energy when the detector is exposed to a pulsed-laser beam of fixed wavelength, should allow measurements of small numbers of incoming photons. However, such measurements are 
limited due to electronic noise in our setup that is at the level of 95~mV. The impact of the noise on the signal amplitude mean is shown in Fig.~\ref{fig:2}~(b). In Fig.~\ref{fig:4} the peak-to-peak voltage amplitude for an output signal spectrum averaged over 1000 events, is shown. The calorimetric mode is sensitive to a number of absorbed photons as low as 15.

\begin{figure}[h!tbp]
\centering
    \includegraphics[width=.48\textwidth]{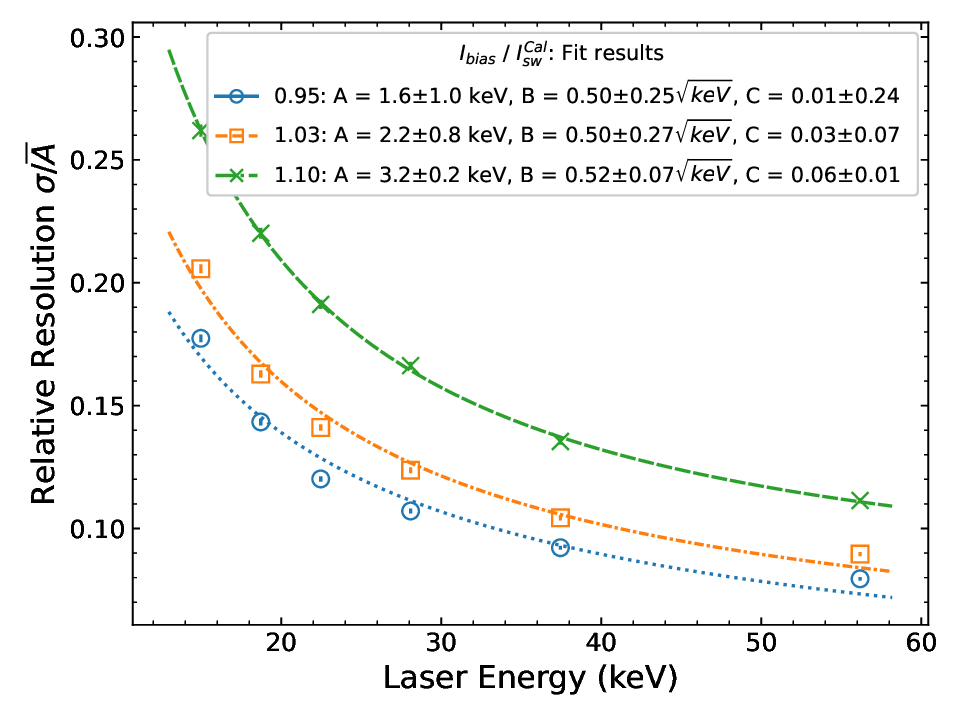}
\caption{SNSPD energy resolution in the calorimetric mode for 3 sample bias currents. In the $x$-axis, energy corresponds to the attenuated laser energy, but it does not include the sensor photon absorption efficiency. The results of fits using a resolution model discussed in the text are reported, where \textit{A} represents the noise term, \textit{B} the stochastic term, and \textit{C} the constant term of the resolution model.
\label{fig:5}
}
\end{figure}

The fractional energy resolution defined as the ratio of the standard deviation of a Gaussian fit of the signal amplitude distribution divided by the amplitude mean, $\sigma/\overline{A}$,
is presented in Fig.~\ref{fig:5} for different laser energies without considering the photon absorption efficiency and for a range of bias currents.
%where $\sigma$ was the standard deviation of the signal %amplitude (\Amp) distribution, and the relative resolution %was defined as $\sigma/\overline{A}$. 
%We observeFor instance, at 10 keV, the energy resolution was around 20\%. 
The fractional resolution is modeled using the expression $\sqrt{((A/E)^2+(B/\sqrt{E})^2+C^2)}$, where \textit{A} corresponds to the noise term, \textit{B} the stochastic term, \textit{C} the constant term, and \textit{E} the laser energy. Fits of the data with this model for three different \Ibias{} are also shown in Fig.~\ref{fig:5}. For bias currents close to the switching current where the amplitude is maximum, we observe a resolution better than 10\% at high energies, while for low energies where the noise contribution dominates, the 0.95 \Ibias{} selection gives the best resolution. The large contribution from noise in these measurements suggests that improving the signal-to-noise ratio in the electronics readout system is a key to optimizing the performance of dual-mode SNSPDs. The average number of absorbed photons \textit{N} can be derived from the fitted stochastic term \textit{B} using the equation $N = E/B^{2}$. This calculation yields an estimated photon absorption efficiency of approximately 1\%. The sources of uncertainty in this estimation include a 3\% uncertainty from the power meter measurements, a 20\% uncertainty due to fitting errors, and a 50\% uncertainty associated with attenuation measurements.

An additional difference between the two operating modes is observed in the effect of latching.
Latching occurs when the time needed for the bias current to return from the load to the SNSPD is shorter than the SNSPD self-reset time, preventing the SNSPD from returning to its original operating state~\cite{annunziata2010}. In the Geiger mode, we observe latching above 0.81 \IswSC, which limits the ability of the detector to operate with a bias current in the saturated internal efficiency regime. The large effective cross-section of our current nanowire design, which results in a smaller kinetic inductance, likely leads to this latching behaviour. In contrast, in the calorimetric mode, latching does not occur for any \Ibias{} (above or below \IswCal). This demonstrates the stable operation of the SNSPD in the calorimetric mode, even when the current return time is shorter than in the Geiger mode. 

In our high-flux measurements the timing jitter for a triggered event is determined by a large number of absorbed photons per laser pulse.
The timing jitter was measured to be approximately 108~ps for both Geiger and calorimetric modes using a Time-Correlated Single-Photon Counting module with a 4~ps timing resolution. It is important to note that the contributions of the 100~ps pulse width, the laser jitter and the electronic noise have not been subtracted from the timing jitter measurement, so the internal timing jitter of the SNSPD is expected to be less than 108~ps. Furthermore, the DCR is low in both modes, approximately 1 count per minute in the Geiger mode, while no dark counts are observed in the calorimetric mode.

The calorimetric behaviour of the SNSPD is analogous to the mechanism observed in transition edge sensors (TES)~\cite{ullom2015}, where the resistance of the superconducting nanowire changes due to variations in temperature and current density from photon absorption. However, there are several performance differences between the calorimetric SNSPD and other cryogenic calorimeters, summarized in Table~\ref{tab:compare}. The timing capability of the calorimetric SNSPD is a significant advantage, offering fast \tauf{} for quick recovery times of the detection system and small timing jitter. However, the energy resolution of the calorimetric SNSPD developed in this study is still electronic-noise limited. 

\begin{table}
\caption{\label{tab:compare}A comparison of the SNSPD in the calorimetric mode presented in this work with state-of-the-art cryogenics bolometers.}
\begin{ruledtabular}
\begin{tabular}{lcrr}
        Detectors &  $\sigma / E$ & \tauf & Timing Jitter\\
        \hline
        TES~\cite{morgan2019,eisaman2011} & 0.06\% (@1 keV) & 87 \us & 10-100 ns\\
        MKID~\cite{swimmer2023,delucia2023} & 1.8\% (@3 eV) & 32 \us & - \\
        NTD-Ge~\cite{cupid,bandler2000} & 0.1\% (@6 keV) & 7 ms & - \\
        Calorimetric SNSPD & <6\% const. term & 560 ps & < 108 ps \\
\end{tabular}
\end{ruledtabular}
\end{table}

The primary challenges for the calorimetric mode include demonstrating single-photon sensitivity with high efficiency and obtaining high-energy resolution for wavelength spectroscopy at the eV level, which would enable fast spectroscopy. Currently, a major issue in our system is external electrical noise, which could be mitigated by employing low-noise cryogenic amplifiers and improving the cryogenic system to be noise-free. Future work should also explore the effects of nanowire geometry. In this study our goal has been to explore the dual-mode operation for a range of temperatures. For this reason, we use a 15~nm thick 200~nm wide NbN nanowire that displays higher \Tc. Such nanowire dimensions simplify our setup in the available cryogenic system.
%, which operates with a minimum temperature of 4 K. 
In this particular nanowire design, the wire is thicker and wider compared to those used in fabricating high-efficiency SNSPDs. Exploring thinner and narrower nanowire geometries is expected to enhance sensitivity. In addition, incorporating optical structures such as mirrors or cavities will further optimize the detector performance.

\section{conclusions}
A dual-mode SNSPD operating in both the conventional Geiger mode and the calorimetric mode is demonstrated. The SNSPD is tested and readout in an external-trigger mode using a pulsed laser emitting photons in the visible range.
Depending on the application,
by tuning the sample temperature and bias current using the same readout system, the SNSPD can readily switch between the Geiger mode for event counting and calorimetric mode for sensitivity to the energy of a monochromatic photon pulse.
This dual-mode capability is enhanced by a low DCR, stable operation in the calorimetric mode, and a relatively simple fabrication process.  At high energies an energy resolution better than 10\% is observed, although at lower energies the resolution performance is limited due to electronic noise. The fast timing capability of these SNSPDs and the dual-mode operation show potential for high-speed spectroscopy and other applications requiring precise timing measurements.

%The advancements detailed in this work, particularly the SNSPD's operation in both geiger and calorimetric modes, offer promising solutions to meet stringent requirements in future applications. 

\section*{SUPPLEMENTARY MATERIAL}
See the \Sup{} for detailed information of the resistance-current diagram and experimental setups.

\begin{acknowledgments}
We wish to acknowledge the support of the NTU-Saclay Joint Laboratory for Infrared Quantum single photon SENSors (IQSENS) grant. This work has been supported in part by the LabEx FOCUS grant, ANR-11-LABX-0013. We also acknowledge financial support from the National Science and Technology Council, Taiwan (Grant NSTC-113-2112-M-001-014), and Academia Sinica of Taiwan (AS-TP-113-M02). The authors would like to thank Stefanos Marnieros and Matias Rodrigues for preparing and establishing the electric contacts on certain SNSPDs.
NbN film \Tc{} measurements where performed using a SQUID at CCMS, NTU, by Li-Min Wang and his group.
\end{acknowledgments}

\section*{AUTHOR DECLARATIONS}
\subsection*{Conflict of Interest}
The authors have no conflicts to disclose.

%\subsection*{Author Contributions}
%\textbf{Hsin-Yeh Wu}: Conceptualization (lead); writing – original draft (equal); Data Curation (lead); Investigation (lead); formal analysis (lead); writing – review and editing (equal). \textbf{Marc Besançon}: Project Administration (equal). \textbf{Jia-Wern Chen}: Resources (equal). \textbf{Pisin Chen}: Project Administration (equal). \textbf{Jean-François Glicenstein}: Project Administration (equal). \textbf{Shu-Xiao Liu}: Investigation (supporting); formal analysis (supporting). \textbf{Yu-Jung Lu}: Funding Acquisition (equal); Supervision (supporting). \textbf{Xavier-François Navick}: Conceptualization (supporting); writing – review and editing (equal). \textbf{Stathes Paganis}: Conceptualization (supporting); Writing – original draft (equal); Supervision (lead); formal analysis (supporting); Writing – review and editing (equal); Funding Acquisition (equal). \textbf{Boris Tuchming}: Conceptualization (supporting); formal analysis (supporting); Writing – review and editing (equal). \textbf{Dimitra Tsionou}: Writing – review and editing (equal). \textbf{Feng-Yang Tsai}: Resources (equal); Investigaion (supporting); formal analysis (supporting)

\section*{Data Availability Statement}
The data that support the findings of this study are available from the corresponding author upon request.

\nocite{*}
\bibliography{DualModeSNSPD}% Produces the bibliography via BibTeX.

\end{document}